\documentclass[prl,preprint,showpacs,preprintnumbers,amsmath,amssymb,superscriptaddress]{revtex4}

\usepackage{graphicx}
\usepackage{dcolumn}
\usepackage{bm}

\begin{document}

\preprint{APS/123-QED}

\title{Mechanism of enhanced optical second-harmonic generation in the conducting pyrochlore-type Pb$_{2}$Ir$_{2}$O$_{7-x}$ oxide compound}

\author{Yasuyuki Hirata}
\affiliation{
Institute for Solid State Physics, The University of Tokyo, Kashiwanoha 5-1-5, Kashiwa, Chiba 277-8581, Japan
}

\author{Makoto Nakajima}
\affiliation{
Graduate School of Science, Chiba University, Chiba 263-8522, Japan
}

\author{Yusuke Nomura}
\affiliation{
Department of Applied Physics, The University of Tokyo, Hongo, Bunkyo-ku, Tokyo 113-8656, Japan
}

\author{Hiroyuki Tajima}
\affiliation{
Institute for Solid State Physics, The University of Tokyo, Kashiwanoha 5-1-5, Kashiwa, Chiba 277-8581, Japan
}

\author{Yoshitaka Matsushita}
\affiliation{
Materials Analysis Station, National Institute for Materials Science, 1-2-1 Sengen, Tsukuba, Ibaraki 305-0047, Japan
}

\author{Keiko Asoh}
\author{Yoko Kiuchi}
\affiliation{
Institute for Solid State Physics, The University of Tokyo, Kashiwanoha 5-1-5, Kashiwa, Chiba 277-8581, Japan
}

\author{Adolfo G. Eguiluz}
\affiliation{
Department of Physics and Astronomy, The University of Tennessee, Knoxville, Tennessee 37996, USA
}

\author{Ryotaro Arita}
\affiliation{
Department of Applied Physics, The University of Tokyo, Hongo, Bunkyo-ku, Tokyo 113-8656, Japan
}

\author{Tohru Suemoto}
\affiliation{
Institute for Solid State Physics, The University of Tokyo, Kashiwanoha 5-1-5, Kashiwa, Chiba 277-8581, Japan
}

\author{Kenya Ohgushi}
\affiliation{
Institute for Solid State Physics, The University of Tokyo, Kashiwanoha 5-1-5, Kashiwa, Chiba 277-8581, Japan
}


\begin{abstract}
The structural, electronic, and optical properties of pyrochlore-type Pb$_{2}$Ir$_{2}$O$_{6}$O'$_{0.55}$, which is a metal without spatial inversion symmetry at room temperature, were investigated. Structural analysis revealed that the structural distortion relevant to the breakdown of the inversion symmetry is dominated by the Pb-O' network but is very small in the Ir-O network. At the same time, gigantic second-harmonic generation signals were observed, which can only occur if the local environment of the Ir 5$d$ electrons features broken inversion symmetry. First-principles electronic structure calculations reveal that the underlying mechanism for this phenomenon is the induction of the noncentrosymmetricity in the Ir 5$d$ bands by the strong hybridization with O' 2$p$ orbitals. Our results stimulate theoretical study of inversion-broken iridates, where exotic quantum states such as a topological insulator and Dirac semimetal are anticipated.
\end{abstract}


\maketitle

Symmetry is a critical factor that controls the physical properties of a solid. Spatial inversion symmetry is one of the most important symmetries. The most common and well-studied systems without inversion symmetry are the ferroelectrics, where macroscopic polarization appears in an insulating state. Recently, conductive materials without inversion symmetry, which are known as "noncentrosymmetric metals" or "polar metals", have also attracted interest\cite{edelstein1, batio3}. In contrast to ferroelectrics, noncentrosymmetric metals do not exhibit macroscopic polarization due to screening by conducting electrons; instead the state is characterized by a higher-rank tensor ({\it e.g.} piezo-electric tensor). The breakdown of inversion symmetry is considered to influence the transport properties. For example, it is theoretically predicted that the inverse Faraday effect can be induced by Rashba interaction.\cite{edelstein1,edelstein2,edelstein3} However, there have been few experimental studies, because noncentrosymmetric metals are rare.

\begin{figure}
\includegraphics[width=16cm]{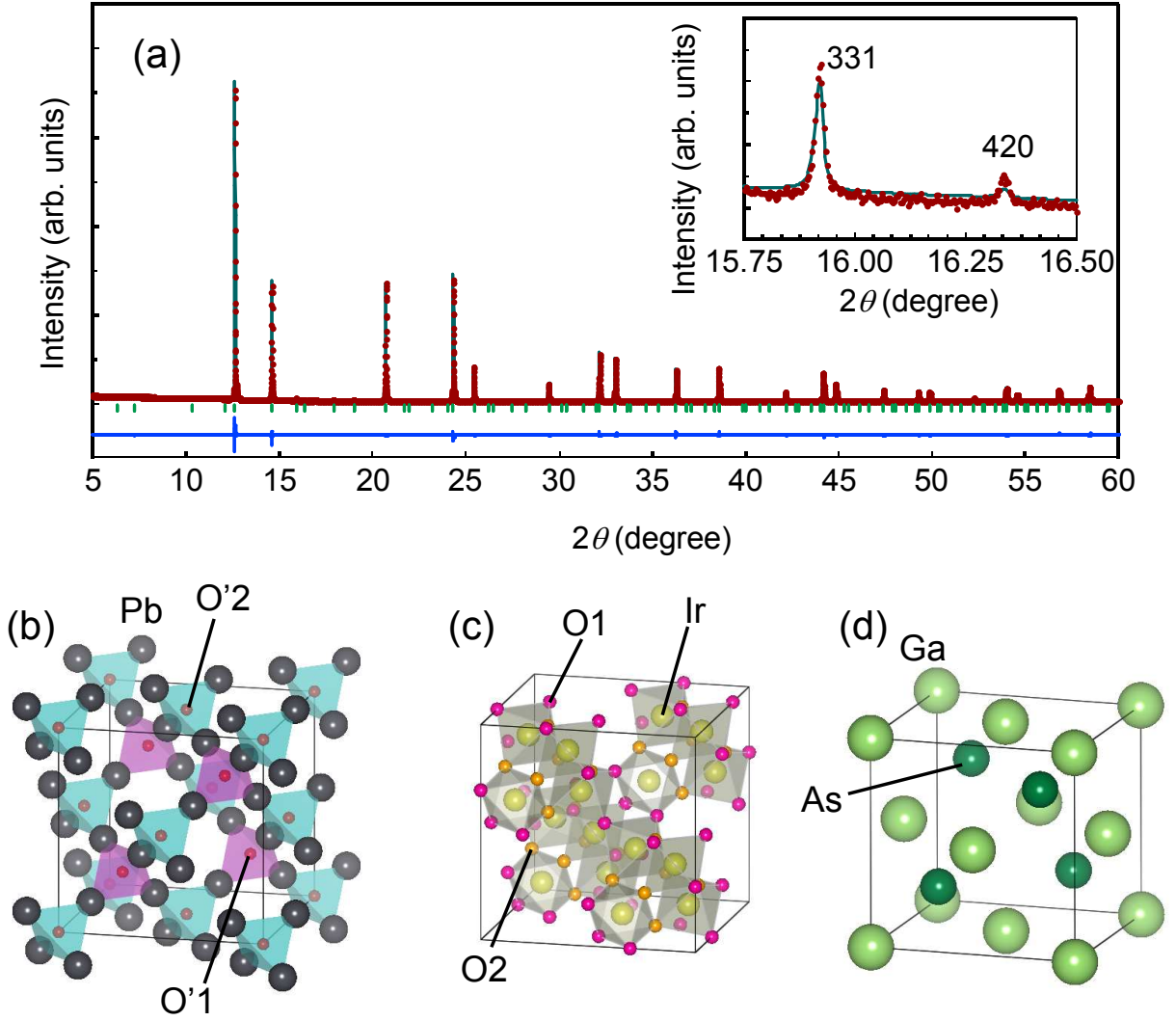}%
\caption{\label{fig:1}(a) Powder X-ray diffraction pattern of Pb$_{2}$Ir$_{2}$O$_{7-x}$. The points denote the experimental data and the solid line represents the fitting curve. The ticks indicate the allowed reflections in $F\bar{4}3m$, and the line below the ticks represents the residual error. The inset shows a closeup of the 3 3 1 and 4 2 0 reflections. (b,c) Crystal structure of Pb$_{2}$Ir$_{2}$O$_{7-x}$, which is formed from the interpenetration of the (b) Pb$_{2}$O' and (c) Ir$_{2}$O$_{6}$ units. (d) Crystal structure of GaAs.}
\end{figure}

There are several noncentrosymmetric metals in the pyrochlore-type transition-metal oxides $A_{2}B_{2}$O$_{6}$O'. For example, Pb$_{2}$Re$_{2}$O$_{7}$ exhibits a structural phase transition at 295 K from cubic centrosymmetric $Fd\bar{3}m$ to cubic noncentrosymmetric $F\bar{4}3m$.\cite{ohgushi} Another example is Cd$_{2}$Re$_{2}$O$_{7}$, which loses inversion symmetry below 200 K. The low temperature symmetry is tetragonal $I\bar{4}m2$, which is one of the subgroups of $F\bar{4}3m$, as revealed by second harmonic generation (SHG) measurements.\cite{petersen} The pyrochlore structure can be divided into two substructures: $A_{2}$O' and $B_{2}$O$_{6}$ units, as shown in Figs.~\ref{fig:1}(b) and \ref{fig:1}(c), respectively. It has been conjectured that the breakdown of inversion symmetry is controlled by the covalency of $A$-O' bonds, whereas the electronic properties are mainly dominated by $d$ electrons in the $B$-O network.\cite{ohgushi} It is essential to unravel the interplay between $A_{2}$O' and $B_{2}$O$_{6}$ units to explore the inversion-related transport phenomena.

Pyrochlore-type Pb$_{2}$Ir$_{2}$O$_{7-x}$ is a unique conductive material in three respects. Firstly, the crystal symmetry is noncentrosymmetric $F\bar{4}3m$, even at room temperature.\cite{kennedy} Secondly, conduction electrons originate from Ir 5$d$ bands, where a strong spin-orbit interaction is present.\cite{ir_so} Thirdly, related compounds $A_{2}$Ir$_{2}$O$_{7}$ ($A$ = Y or rare earth metals) exhibit thermal metal-insulator transitions caused by a strong electron correlation effect.\cite{lio_mit} Despite these intriguing features, there have been no reported studies using a single-crystalline sample.

In this Letter, we report the structural, electronic, and optical properties of single-crystalline Pb$_{2}$Ir$_{2}$O$_{7-x}$ with special attention given to the inversion symmetry. While the structural analysis shows that the inversion symmetry breaking is dominated by Pb-O' bonds, the observed SHG signals, which mainly originate from the noncentrosymmetricity of the Ir 5$d$ band, are strong. We demonstrate that this phenomena is caused by the large hybridization of Ir 5$d$ and O' 2$p$ orbitals, which is unique characteristics of Pb-containing pyrochlore-type oxides.

Single crystals of Pb$_{2}$Ir$_{2}$O$_{7-x}$ with sizes of 2 mm$^{3}$ were grown by the self-flux method. PbO and IrO$_{2}$ powders were mixed in a molar ratio of 9:1, heated to 1250 $^{\circ}$C, and then cooled to 950 $^{\circ}$C at a cooling rate of 5 $^{\circ}$C/h. Scanning electron microscopy/energy dispersive X-ray analysis showed that the content ratio of Pb to Ir was 1.0. The oxygen deficiency was determined to be $x$ = 0.45 by the thermogravimetric/differential thermal analysis method. Resistivity, the Hall coefficient, specific heat, and magnetic susceptibility were measured by using a commercial setup. The synchrotron powder X-ray diffraction data for crushed single crystals was collected at the beamline BL15XU, SPring-8, with wavelength 0.065287 nm. Reflectivity spectra were measured for the (1 1 1) surface polished with Al$_{2}$O$_{3}$ powders. Optical conductivity spectra were obtained by the Kramers-Kronig transformation. The SHG signals were detected using the geometry shown in Fig.~\ref{fig:4}(a). Band structure calculation based on density functional theory\cite{calc1,calc2} with the Elk full-potential linearized augmented plane-wave code\cite{calc3}, which includes the spin-orbit interaction, were performed by using experimentally obtained structural parameters with the $F\bar{4}3m$ crystal symmetry. The muffin tin radii ($R_{\rm MT}$) of 2.0, 2.0, and 1.6 bohr for Pb, Ir and O were used, respectively. The maximum modulus for the reciprocal vectors $K_{\rm max}$ was chosen such that $R_{\rm MT}^{\rm min} K_{\rm max} = 7.0$, where $R_{\rm MT}^{\rm min}$ is the smallest $R_{\rm MT}$ in the system. We employed the Perdew-Burke-Ernzerhof exchange-correlation functional \cite{calc4} and $10\times10\times10$ ${\mathbf k}$-mesh.

\begin{table}
\caption{\label{tab:1} 
Final refined structural parameters of Pb$_{2}$Ir$_{2}$O$_{6.55}$ with noncentrosymmetric space group $F\bar{4}3m$. The lattice constant is $a$ = 1.027149(2) nm. Site occupancy and isotropic atomic displacement parameter are denoted as $g$ and $U$, respectively. The reliability indices \cite{rietveld} of this fit are $R_{\rm wp}$ = 2.45\% and $R_{\rm p}$ = 1.64\%.}
\begin{tabular}{c|lllll}
\hline
 & $g$ & $x$ & $y$ & $z$ & $U$ [nm$^{2}$] \\ \hline
Pb & 1 & 0.87742(5) & 0.87742(5) & 0.87742(5) & 0.000102(2) \\
Ir & 1 & 0.37520(4) & 0.37520(4) & 0.37520(4) & 0.00054(2) \\
O1 & 1 & 0.303(1) & 0 & 0 & 0.00013(1) \\
O2 & 1 & 0.448(1) & 0.25 & 0.25 & 0.00013(1) \\
O'1 & 0.55 & 0.75 & 0.75 & 0.75 & 0.00032(8) \\
O'2 & 0.55 & 0 & 0 & 0 & 0.00032(8) \\ \hline
\end{tabular} 
\end{table}

Figure \ref{fig:1}(a) shows the synchrotron powder X-ray diffraction pattern for Pb$_{2}$Ir$_{2}$O$_{7-x}$. The 4 2 0 reflection, which is allowed in noncentrosymmetric $F\bar{4}3m$ and forbidden in centrosymmetric $Fd\bar{3}m$, is clearly observed [inset of Fig.~\ref{fig:1}(a)]. The 4 2 0 reflection does not vanish, even at 800 $^{\circ}$C (data not shown), which indicates the robustness of the noncentrosymmetric phase. Rietveld analysis was performed using the RIETAN-FP software\cite{rietan} with the assumption that only the O' site can be vacant; thus obtained structural parameters are summarized in Table \ref{tab:1}. The O and O' sites are split into two sites, O1/O2 and O'1/O'2, respectively, which reflects the loss of inversion symmetry in $F\bar{4}3m$. Consequently, two types of chemical bonds appear between a cation and an O$^{2-}$ ion with distinguishable bond length. The splitting ratios between the longer and shorter bond lengths are 0.5\% for the Pb-O bond, 4.1\% for the Pb-O' bond, and 0.2\% for the Ir-O bond. A fairly large bond length splitting for Pb-O' compared with other bonds indicates that the breakdown of the inversion symmetry is dominated by distortion of the Pb-O' bonds. This is most likely due to the strong covalency of the Pb-O' bond.\cite{bi2sn2o7,bi2ti2o7} The breaking of the inversion symmetry is well interpreted by extracting the Pb$_{2}$O' unit, as shown in Fig.~\ref{fig:1}(b). In the centrosymmetric $Fd\bar{3}m$ phase, O'Pb$_{4}$ tetrahedra with equal volume form the diamond structure. When the inversion symmetry is broken, expanded and shrunk tetrahedra are arrayed alternately. The pattern is completely the same as the arrangement of Ga and As atoms in GaAs with the zinc blende structure, which also has noncentrosymmetric $F\bar{4}3m$ symmetry [Fig.~\ref{fig:1}(d)].

\begin{figure}
\includegraphics[width=16cm]{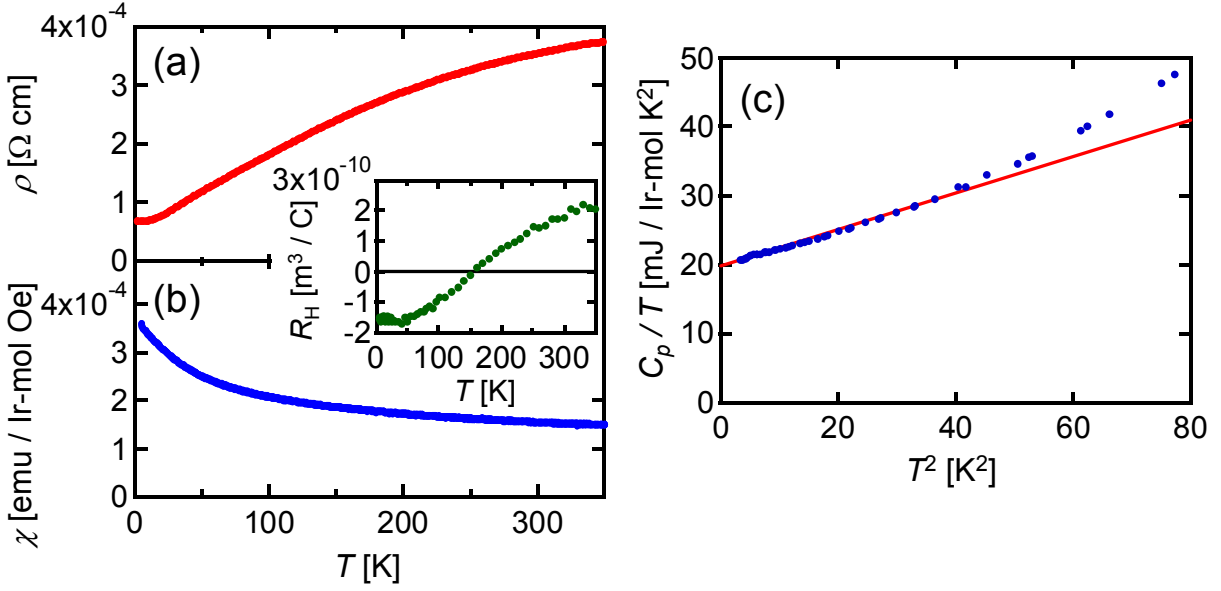}%
\caption{\label{fig:2} (a, b) Temperature ($T$) dependence of (a) resistivity ($\rho$), and (b) magnetic susceptibility ($\chi$) for Pb$_{2}$Ir$_{2}$O$_{7-x}$. The inset shows the Hall coefficient ($R_{\rm H}$) dependence. (c) Temperature dependence of specific heat ($C_{p}$) divided by $T$ for Pb$_{2}$Ir$_{2}$O$_{7-x}$. The solid line represents the fitting results for the range between $T$ = 1.8 and 6.4 K with the relation $C_{p} = \gamma T+\beta T^{3}$.}
\end{figure}

We move to consider electronic properties originating from conduction electrons in the Ir$_{2}$O$_{6}$ unit. The temperature dependence of resistivity $\rho$, magnetic susceptibility $\chi$, the Hall coefficient $R_{{\rm H}}$, and specific heat $C_{p}$ are shown in Fig.~\ref{fig:2}. The resistivity indicates a metallic nature over the entire temperature range measured; mobile carriers consist of both holes and electrons, as evidenced by the zero-crossing feature of $R_{{\rm H}}$. The magnetic susceptibility is represented as a sum of the Pauli-paramagnetic component $\chi_{0}$ = 2.33$\times$10$^{-4}$ emu/Ir-mol$\cdot$Oe and the Curie-Weiss-like component $C/(T-\theta_{\rm W})$ with $C$ = 2.83$\times$10$^{-2}$ emu$\cdot$K/Ir-mol$\cdot$Oe and $\theta_{\rm W}$ = $-$53.4 K; the latter term possibly originates from crystal imperfections, as observed in Pb$_{2}$Ru$_{2}$O$_{6.5}$ and Bi$_{2}$Ru$_{2}$O$_{7}$.\cite{pb2ru2o7,bi2ru2o7} The specific heat is fitted in the range between $T$ = 1.8 and 6.4 K by the relation $C_{p} = \gamma T+\beta T^{3}$, where the former and latter terms denote the electronic and lattice contributions, respectively. We obtained $\gamma$ = 19.7 mJ/Ir-mol$\cdot$K$^{2}$ and $\beta$ = 0.27 mJ/Ir-mol$\cdot$K$^{4}$. The Wilson ratio $R_{\rm W}$ = $\pi^{2}k_{B}^{2}\chi_{s}/3\mu_{B}^{2}\gamma$ is calculated to be 0.85 with the assumption of $\chi_{s} = \chi_{0}$, indicating a weak correlation effect. The measurable discrepancy of $C_{p}$ from the relation $C_{p} = \gamma T+\beta T^{3}$ [Fig. 2(c)] is also observed in Pb$_{2}$Ru$_{2}$O$_{6.5}$ and Bi$_{2}$Ru$_{2}$O$_{7}$,\cite{bi2ru2o7} and likely originates from the low-energy optical phonons associated with heavy Pb and Bi atoms.

\begin{figure}
\includegraphics[width=12cm]{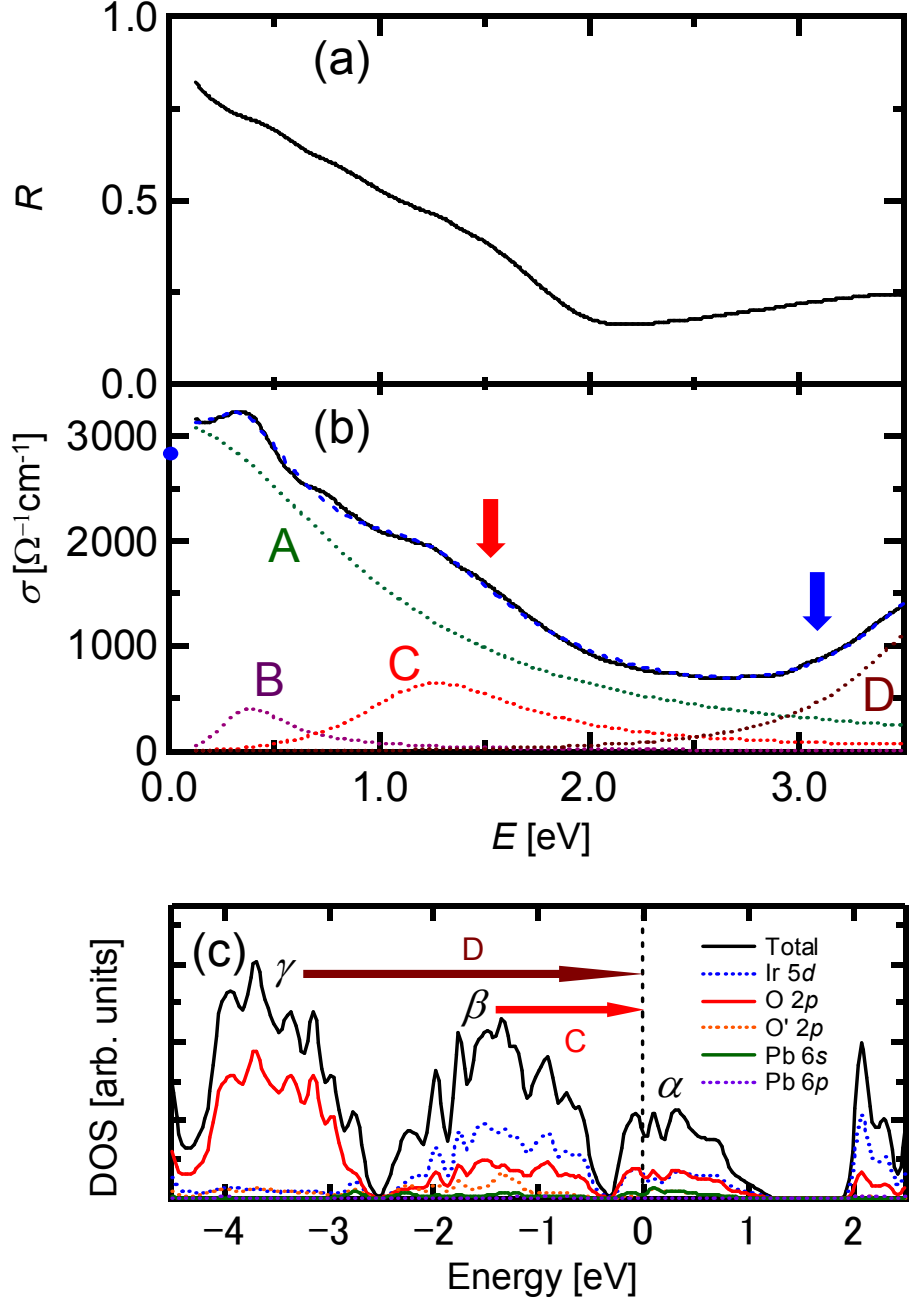}%
\caption{\label{fig:3}(a) Reflectivity ($R$) spectra as a function of photon energy ($E$) for Pb$_{2}$Ir$_{2}$O$_{7-x}$ at room temperature. (b) Optical conductivity ($\sigma$) of Pb$_{2}$Ir$_{2}$O$_{7-x}$ at room temperature. The closed circle marked at $E$ = 0 eV indicates the dc conductivity obtained from the resistivity measurement. The dashed line shows the full Drude-Lorentz fit. Dotted lines labeled as A, B, C and D represent the extracted Drude-Lorentz components. The arrows indicate the energy of the incident (1.55 eV) and generated (3.10 eV) beams of the SHG measurement. (c) Total density of states (DOS), and partial density of states of Ir 5$d$, O 2$p$, O' 2$p$, Pb 6$s$ orbitals, and Pb 6$p$ orbitals, which are obtained from the first-principles calculation.}
\end{figure}

Reflectivity and optical conductivity spectra are shown in Figs.~\ref{fig:3}(a) and (b), respectively. In the optical conductivity spectrum, besides the Drude component with the plasma frequency $\omega_{p}\sim$2 eV, two broad structures at 0.4 and 1.3 eV are clearly evident. By applying a Drude-Lorentz fitting analysis, four components were extracted [labeled as A, B, C, and D in Figs.~\ref{fig:3}(b)]. These modes can be understood by referring to the band scheme obtained from the first-principles calculation [Fig.~\ref{fig:3}(c)]. Owing to the strong spin-orbit coupling in the 5$d$ transition metal system, Ir 5$d$ $t_{2g}$ orbitals would be reconstructed into the complex orbital states, which are the doubly-degenerate $J_{\rm eff}$ = 1/2 and fourfold-degenerate $J_{\rm eff}$ = 3/2 states, if there were no crystal field.\cite{kim1,kim2,cairo3,moon,arita} Under the ligand field comparable to the spin-orbit interaction, $J_{\rm eff}$ = 1/2 states and $J_{\rm eff}$ = 3/2 states are hybridized with O/O' 2$p$ orbitals and form the $\alpha$ and $\beta$ bands near the Fermi level. The $\gamma$ band mostly consists of O 2$p$ orbitals. Therefore, we can assign the mode observed in the optical conductivity spectra as follows: the Drude term (A) represents the plasma excitation within the $\alpha$ band; the 1.3 eV mode (C) corresponds to the excitation from the $\beta$ band to the $\alpha$ band; and the high-energy mode (D) is the charge transfer excitation from the O 2$p$ $(\gamma)$ band to the $\alpha$ band. The origin of 0.4 eV (B) mode is still unclear; however, we speculate that the mode is the intraband transition among the $\alpha$ band and/or the excitation associated with the self-doped hole carriers in the $\beta$ band, which are stabilized by the local Hund coupling.

\begin{figure}
\includegraphics[width=16cm]{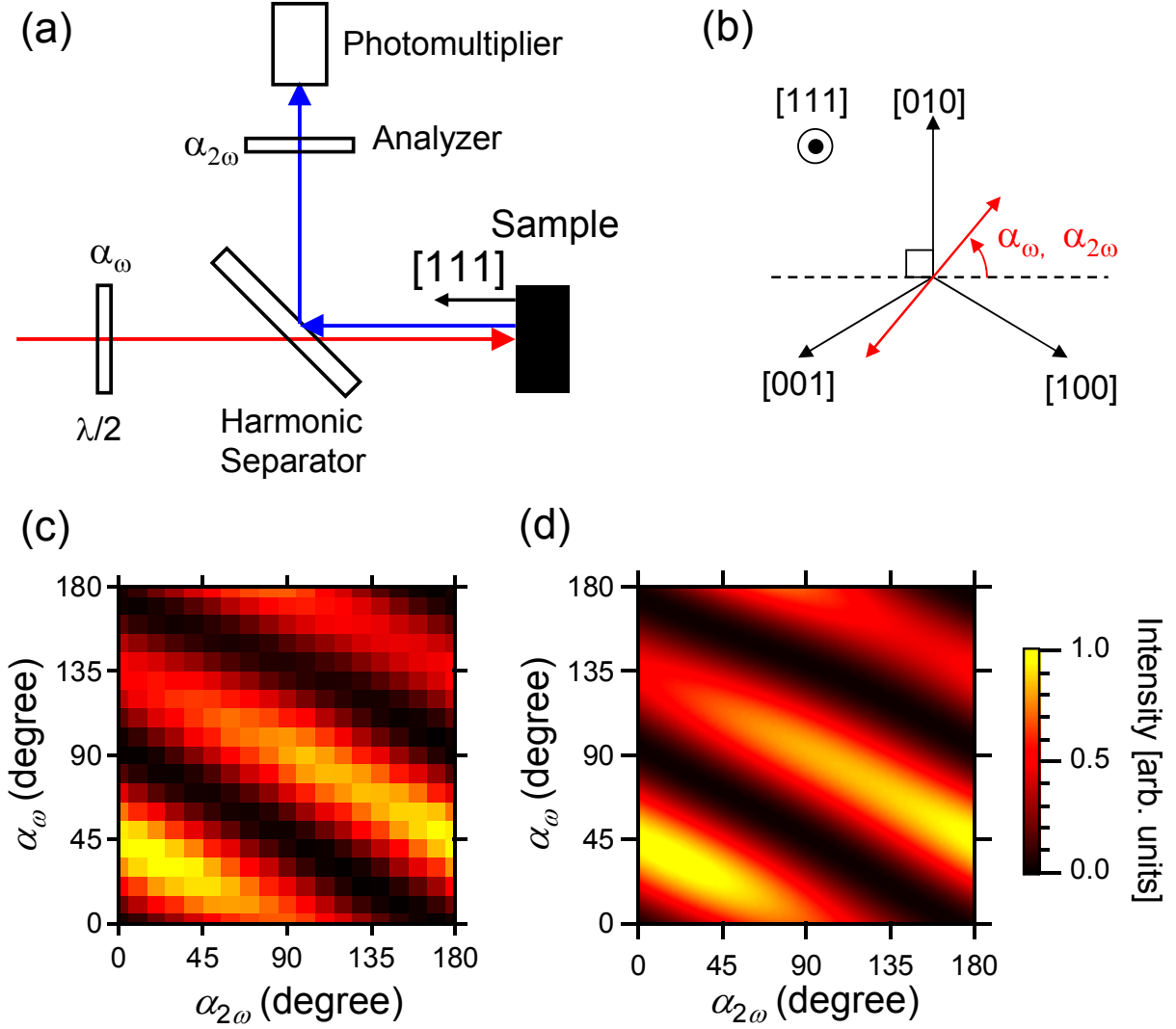}%
\caption{\label{fig:4}(a) Experimental geometry of the SHG measurement. Incident pulsed light with a wavelength of 800 nm (1.55 eV) is generated using a Ti:sapphire laser. The pulse duration was 100 fs at a frequency of 82 MHz, with a power of 30 mW in a spot size of 100 $\mu$m. The incident beam vertically directed on the (1 1 1) surface generates a SHG beam with a wavelength of 400 nm (3.10 eV), which is extracted by a harmonic separator toward the photomultiplier detector. The half-wave plate and the crystal polarizer control the polarization of the incident and generated beams, respectively. (b) Relationship between the crystal axis of Pb$_{2}$Ir$_{2}$O$_{7-x}$ and the polarization angle of incident/generated light $\alpha_{\omega}$/$\alpha_{2\omega}$ on the (1 1 1) surface. (c) Polarization dependence of the observed SHG signal for Pb$_{2}$Ir$_{2}$O$_{7-x}$. (d) Simulated bulk SHG signal for an $F\bar{4}3m$ crystal of which the [1 1 1] axis is tilted by 5$^{\circ}$ from the surface normal.}
\end{figure}

In order to investigate the impact of broken inversion symmetry on the electronic states, we performed an SHG measurement, summarized in Fig. \ref{fig:4}. The photon energy of the incident 800 nm beam ($E$ = 1.55 eV) and generated 400 nm beam (2$E$ = 3.10 eV) are represented by arrows in Fig.~\ref{fig:3}(b). Since the optical responses of both energies are dominated by electrons within the Ir 5$d$ and O 2$p$ bands, the SHG signals mainly reflect the noncentrosymmetricity of the Ir 5$d$ and O 2$p$ bands. The observed polarization dependence of the SHG signals shown in Fig.~\ref{fig:4}(c) coincides well with the simulated bulk SHG signal, where the [1 1 1] axis is tilted by 5$^{\circ}$ from the surface normal[Fig.~\ref{fig:4}(d)].\cite{shg} This ensures that the observed signals come from the bulk SHG response.\cite{shg_comparison} The SHG signal of a GaAs single crystal was also measured under the same experimental conditions and the intensity was compared with that from Pb$_{2}$Ir$_{2}$O$_{7-x}$; the intensity ratio is $I$(Pb$_{2}$Ir$_{2}$O$_{7-x}$)/$I$(GaAs) = 0.7. This result is surprising, because the breakdown of the inversion symmetry in Pb$_{2}$Ir$_{2}$O$_{7-x}$ is caused merely by atomic displacement in the Pb$_{2}$O' unit, while that of GaAs originates from the alternating array of two distinguishable atoms. Furthermore, the SHG indicates the noncentrosymmetricity of the Ir 5$d$ and O 2$p$ bands, which has seemingly no direct connection with the Pb$_{2}$O' unit.

Our first-principles electronic structure calculations unravel the underlying mechanism of this unexpectedly large SHG signal. The partial density of states (PDOS) of $\beta$ bands [Fig. \ref{fig:3}(c)] points to the relatively strong hybridization of O' 2$p$ orbitals with Ir 5$d$ orbitals in spite of a long spatial distance (the pathway being O'-Pb-O-Ir). When we divide the PDOS by the number of atoms in a formula unit (6 for O and 1 for O'), we notice that the contribution of the O' 2$p$ orbital is even larger than that of the O 2$p$ orbital. This indicates that even though Pb$_{2}$O' and Ir$_{2}$O$_{6}$ units are separated structurally, they are tightly coupled with each other electronically. Such a large hybridization between $B$ $d$ and O' 2$p$ orbitals in the calculated band structure is rarely reported in other pyrochlores with Y and rare earth atoms as $A$ site; an exceptional example is Tl$_{2}$Ru$_{2}$O$_{7}$.\cite{ishii_calc,y2ir2o7_calc} The characteristic feature common to Pb$_{2}$Ir$_{2}$O$_{7-x}$ and Tl$_{2}$Ru$_{2}$O$_{7}$ is that the 6$s$ orbital is located at close to the Fermi energy ($-$9 eV for $A$=Pb and 2 eV for $A$=Tl) and is strongly hybridizing with O 2$p$ orbitals. Consequently, Pb$_{2}$O' and Ir$_{2}$O$_{6}$ units are electronically bridged, leading to the strong hybridization of Ir 5$d$ and O' 2$p$ orbitals.

Once we recognize that the strong hybridization between Ir 5$d$ and O' 2$p$ orbitals induces the noncentrosymmetricity in the Ir 5$d$ bands, the enhanced SHG signal can be explained as follows. Since the incident and generated photon energies coincide with the C and D bands in the optical conductivity spectrum, respectively, we can approximately express the SHG susceptibilty tensor of as
\begin{equation*}
\chi^{(2)}_{jkl}(2\omega; \omega, \omega)\propto \frac{\langle\alpha|\mu^{j}|\gamma\rangle\langle\gamma|\mu^{k}|\beta\rangle\langle\beta|\mu^{l}|\alpha\rangle}{(E_{\alpha\gamma}-2\hbar\omega-\Gamma_{\alpha\gamma})(E_{\alpha\beta}-\hbar\omega-\Gamma_{\alpha\beta}))},
\end{equation*}
where $\mu$ denotes the dipole operator, $\Gamma_{\alpha\gamma}$ and $\Gamma_{\alpha\beta}$ denote damping factors, $E_{\alpha\beta}$ and $E_{\alpha\gamma}$ denote the excitation energy from the $\beta$ to $\alpha$ band, and from the $\gamma$ to $\alpha$ band, respectively.\cite{boyd} The matrix element $\langle\beta|\mu|\alpha\rangle$ stands for the $d$-$d$ transition and hence is originally dipole-forbidden. However, the hybridization of Ir 5$d$ orbitals and O' 2$p$ orbitals, the latter of which are sensitive to the noncentrosymmetricity of the Pb$_{2}$O' unit, revives this virtual excitation. Then, $\chi^{(2)}_{jkl}$ becomes finite, which well explains the large SHG signals.

Recent theoretical studies on the pyrochlore-type iridates have revealed that the ground state exhibits a wide variety, ranging from a correlated insulator, a topological insulator, to a Dirac semi-metal, as a consequence of the interplay between the electron correlation effect, spin-orbit coupling, and band filling.\cite{yio_tbi,yio_wsm} However, the position of Pb$_{2}$Ir$_{2}$O$_{7-x}$ among the global phase diagram is still unclear, because all the proposed theories suppose the inversion symmetry. Further theoretical studies on inversion-broken systems are therefore necessary to further understand the electronic behavior of pyrochlore-type iridates. 

In conclusion, the structural, electronic, and optical properties of pyrochlore-type Pb$_{2}$Ir$_{2}$O$_{7-x}$ were investigated. Structural analysis indicates that breakdown of the inversion symmetry of the lattice sector is dominated by the Pb$_{2}$O' unit. Nevertheless, gigantic bulk SHG signals comparable to those of GaAs reveal the strong noncentrosymmetricity of the electrons in the Ir 5$d$ band. We have shown that the electronic non-centrosymmetricity originates from the large hybridization of O' 2$p$ orbital with Ir 5d band, which is mediated by the closeness of energy level of Pb 6$s$ and O 2$p$ orbitals.

We are grateful to Y. Ueda and T. Arima for helpful discussions and to M. Isobe, J. Yamaura, H. Ohsumi, S. Takeshita, and S. Uchida for experimental support. This work was supported by Special Coordination Funds for Promoting Science and Technology, Promotion of Environmental Improvement for Independence of Young Researchers, a Grant-in-Aid for Young Scientists (A) (No.23686012), (B) (No. 20740211), and (C) (No. 21540330). A.G.E. acknowledges support by NSF Grant OCI-0904972.


\begin{thebibliography}{99}
\bibitem{shg}
The nonlinear susceptibility $\chi^{(2)}_{jkl}$ for SHG is defined as
\begin{equation}
P_{j}(2\omega) = \chi^{(2)}_{jkl}(2\omega;\omega,\omega)E_{k}(\omega)E_{l}(\omega),
\end{equation}
where $P_{j}$ denotes the polarization, and $E_{k}$ and $E_{l}$ denote the electric field.\cite{boyd} The crystal symmetry $F\bar{4}3m$ reduces independent $\chi^{(2)}_{jkl}$; $\chi^{(2)}_{123} = \chi^{(2)}_{213} = \chi^{(2)}_{312} = \chi^{(2)}_{132} = \chi^{(2)}_{231} = \chi^{(2)}_{321} = \chi^{(2)}$, and all the other elements are zero. When incident light is irradiated on the (1 1 1) surface, the generated polarization is expressed as
\begin{eqnarray}
\left(\begin{array}{c} P_{x} \\ P_{y} \end{array}\right) = \chi^{(2)}\left(\begin{array}{c} 2E_{x}E_{y} \\ E_{x}^{2}-E_{y}^{2} \end{array}\right),
\end{eqnarray}
using the coordinate axis shown in Fig.~\ref{fig:4}(b).
\bibitem{shg_comparison}
The observed SHG signals correspond to the $A_{2u}$ order parameter $\tau$ in Ref.~\onlinecite{petersen}. In the SHG measurement of Cd$_{2}$Re$_{2}$O$_{7}$ the dominant SHG signals come from the $E_{u}^{(2)} $ order parameter $\eta_{2}$, which corresponds to tetragonal distortion of the crystal, and $\tau$ is one order of magnitude smaller than $\eta_{2}$, because $\tau$ is a higher order parameter. In contrast, our sample has no tetragonal distortion, so that the SHG signals from $\tau$ become dominant.
\end{thebibliography}
\end{document}